\documentclass[aps,prb,preprint,groupedaddress]{revtex4}
\usepackage{graphicx} 
\usepackage{dcolumn}% Align table columns on decimal point
\usepackage{bm}% bold math

\begin{document}

\title{ Strong shape dependence of the Morin transition in  $\alpha$-Fe$_{2}$O$_{3}$ single-crystalline nanostructures } 
\author{Jun Wang$^{1,a}$, Victor Aguilar$^{2}$, Le Li$^{1}$, Fa-gen Li$^{1}$, Wen-zhong Wang$^{3}$, and 
Guo-meng Zhao$^{1,2,b}$} 
\affiliation{
$^{1}$Department of Physics, Faculty of Science, Ningbo
University, Ningbo, P. R. China~\\
$^{2}$Department of Physics and Astronomy, 
California State University, Los Angeles, CA 90032, USA~\\
$^{3}$School of Science,  Minzu University of China, Beijing 100081, P. R. China}

\begin{abstract}
  Single-crystalline $\alpha$-Fe$_{2}$O$_{3}$  nanorings (short nanotubes) and nanotubes were 
synthesized by a hydrothermal method. High-resolution transmission electron microscope  and selected-area electron diffraction confirm that the axial directions of both nanorings and nanotubes  are parallel to the crystalline $c$-axis. What is intriguing is that
the Morin transition occurs at about 210 K in the short nanotubes with a mean tube length of about 115~nm and a mean outer diameter of 169~nm while it disappears in the nanotubes with a mean tube length of about 317~nm and a mean outer diameter of 148~nm. Detailed analyses of magnetization data, x-ray diffraction spectra, and room-temperature M\"ossbauer spectra demonstrate that this very strong shape dependence of the Morin transition is intrinsic to hematite.  We can quantitatively explain this intriguing shape dependence  in terms of opposite signs of the surface magnetic anisotropy constants in the surface planes parallel and perpendicular to the  $c$-axis (that is, $K_{s\parallel}$ = -0.37~erg/cm$^{2}$ and $K_{s\perp}$ = 0.42~erg/cm$^{2}$).

\end{abstract}
\maketitle 

\section{Introduction}

Hematite ($\alpha$-Fe$_{2}$O$_{3}$) has a corundum crystal structure and
orders antiferromagnetically below its N\'eel temperature of about 960 K. Bulk hematite exhibits a Morin transition  \cite{Morin}
at about 260 K, below which it is in an antiferromagnetic (AF) phase, where the two antiparallel
sublattice spins are aligned 
along the rhombohedral
[111] axis. Above the Morin transition
temperature $T_{M}$, $\alpha$-Fe$_{2}$O$_{3}$ is in the weak 
ferromagnetic (WF) phase, where the antiparallel spins are slightly canted and lie in 
the basal (111) plane rather than along [111] axis.  The Morin transition is companied by the change of the total magnetic anisotropic constant  from a negative value at $T > T_{M}$ to a positive value at $T < T_{M}$.  Interestingly, this AF-WF transition 
was found to depend on magnetic field. An applied magnetic field parallel to the 
rhombohedral
[111] axis below $T_{M}$ was shown \cite{Besser,Foner,Hirone} to induce the spin-flip transition in the entire 
temperature range below $T_{M}$. The AF-WF transition can also be induced by an applied 
magnetic field  perpendicular to the [111] direction \cite{Flanders}.  The magnetic structure, the Morin 
transition, and  the field dependence of $T_{M}$ were explained
\cite{Dzy58,Flanders} in terms of 
phenomenological thermodynamical potential of Dzyaloshinsky.

In recent years, magnetic nanostructures have attracted
much attention, not only because of their interesting
physical properties but also because of their broad technological applications. Of particular 
interest is a finite-size effect on ferromagnetic/ferrimagnetic  transition temperature.  
Finite-size effects have been  studied in quasi-two-dimensional ultra-thin 
ferromagnetic films \cite{Huang,Li,Elmers,Sch}  and in 
quasi-zero-dimensional ultra-fine ferromagnetic/ferrimagnetic nanoparticles \cite{Tang,Du,ZhaoAPL,ZhaoPRB,ZhaoJAP}. The studies on thin films \cite{Huang,Li,Elmers,Sch}  and recent studies on nanoparticles \cite{ZhaoAPL,ZhaoPRB,ZhaoJAP} have consistently confirmed the finite-size scaling relationship predicted earlier \cite{Fisher}.  
Similarly,  a finite-size effect on the Morin transition temperature was observed in 
nanosized $\alpha$-Fe$_{2}$O$_{3}$ spherical particles \cite{Schr,Gall,Mu,Morr,Bod}. The data show that $T_{M}$ decreases 
with decreasing particle size \cite{Schr,Mu}, similar to the case of ferromagnetic/ferrimagnetic nanoparticles  \cite{ZhaoAPL,ZhaoPRB,ZhaoJAP}. The reduction in the Morin transition temperature was interpreted as due to inherent lattice strain (lattice expansion) of  
nano-crystals~\cite{Schr,Morr}.  It was also shown  \cite{Mu}  that the $T_{M}$  suppression is caused by both strain and the finite-size effect, commonly observed in ferromagnetic/ferrimagnetic materials.  More recently,  Mitra {\em et al.} \cite{Mitra} have found that $T_{M}$ shifts from 251~K for ellipsoidal  to 221~K for rhombohedral nanostructure, which suggests observable shape dependence of the Morin transition.  Here we show that the Morin transition temperature depends very strongly on the shape of nanocrystals: $T_{M}$ shifts from 210~K for the nanorings (short nanotubes with a mean tube length of about 115~nm) to $<$10~K for the nanotubes with a mean tube length of about 317~nm.  The very strong shape dependence of the Morin 
transition is quite intriguing considering the fact that the lattice strains of both nanoring and nanotube crystals are negligibly small, and that the sizes of nanotubes are too large to explain their complete suppression of $T_{M}$ by a finite-size effect. Instead, we can quantitatively explain this intriguing shape dependence  in terms of opposite signs of the surface magnetic anisotropy constants in the surface planes parallel and perpendicular to the  $c$-axis. 

\begin{figure}[htb]
     \vspace{-0.2cm}
    \includegraphics[height=11.0cm]{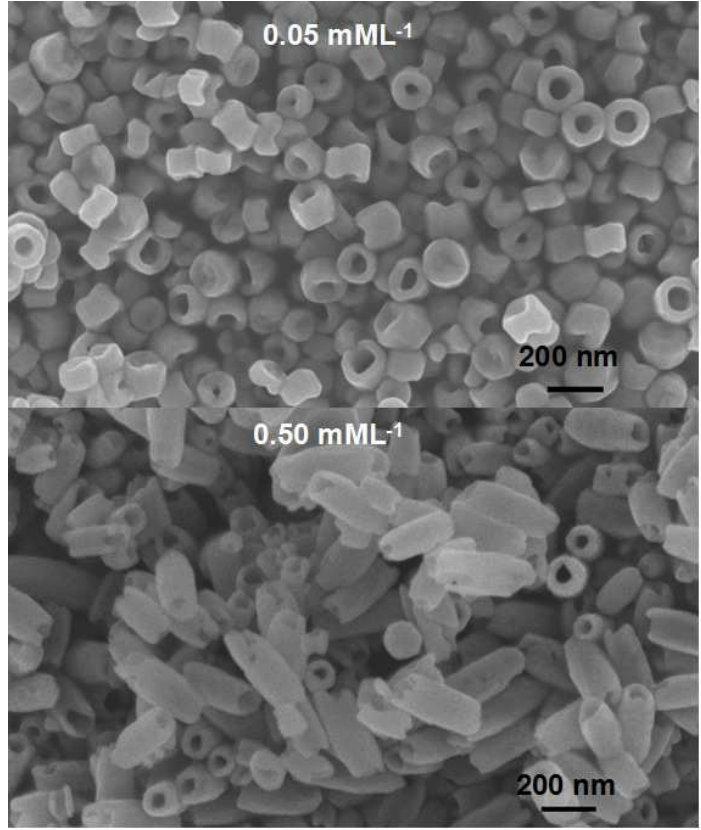}
     \vspace{-0.1cm}
 \caption[~]{Scanning electron microscopic images of the two $\alpha$-Fe$_{2}$O$_{3}$ nanostructures prepared with different phosphate concentrations.  A ring-like morphology (short nanotubes) is seen in the sample prepared with the phosphate concentration of 0.05 mM/L (upper panel) and a tube-like morphology is observed in the sample prepared with the phosphate concentration of 0.50 mM/L (lower panel).  } 
\end{figure} 

\section{ Experimental Results}
 
$\alpha$-Fe$_{2}$O$_{3}$ nanorings were prepared by a hydrothermal method, which is similar to that reported in \cite{Jia}. In the typical 
process, FeCl$_{3}$, NH$_{4}$H$_{2}$PO$_{4}$ (phosphate), and Na$_{2}$SO$_{4}$ were dissolved in deionized 
water with concentrations of 0.002, 0.05 and 0.55 mM/L, respectively.  After vigorous 
stirring for 15 min, the mixture was transferred into a Teflon-lined stainless steel autoclave 
for hydrothermal treatment at 240 $^{\circ}$C for 48 h. While keeping all other experimental 
parameters unchanged, increasing the phosphate concentration from 0.05  to 0.50 mM/L to produce  nanotubes.

\begin{figure}[htb]
     \vspace{-0.2cm}
    \includegraphics[height=12.0cm]{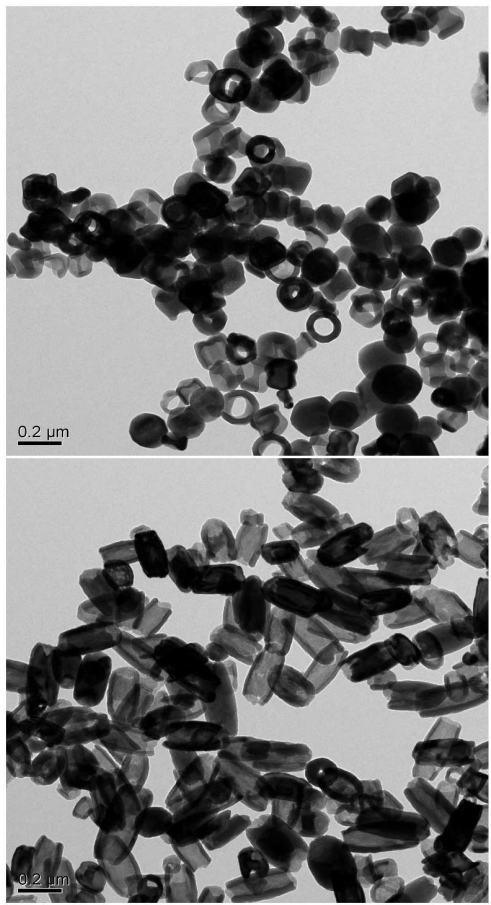}
     \vspace{-0.1cm}
 \caption[~]{Transmission electron microscopic images of the two $\alpha$-Fe$_{2}$O$_{3}$ nanostructures.  The upper panel is the image for the nanorings and  the lower panel is  for the nanotubes.  } 
\end{figure} 

The morphology of the samples was analyzed by field emission scanning electron microscopy
(FE-SEM, SU70, operated at 3 kV) and transmission electron microscopy (JEOL-2010, operated at 200 kV). Figure~1 shows scanning electron microscopic (SEM) images of the two $\alpha$-Fe$_{2}$O$_{3}$ nanostructures prepared with different phosphate concentrations. A ring-like morphology is seen in the sample prepared with the phosphate concentration of 0.05 mM/L (upper panel) and a tube-like morphology is observed in the sample prepared with the phosphate concentration of 0.50 mM/L (lower panel). In fact, these nanorings can be described as short nanotubes with tube lengths shorter than outer diameters.

\begin{figure}[htb]
     \vspace{-0.2cm}
   \leftline{\includegraphics[height=11cm]{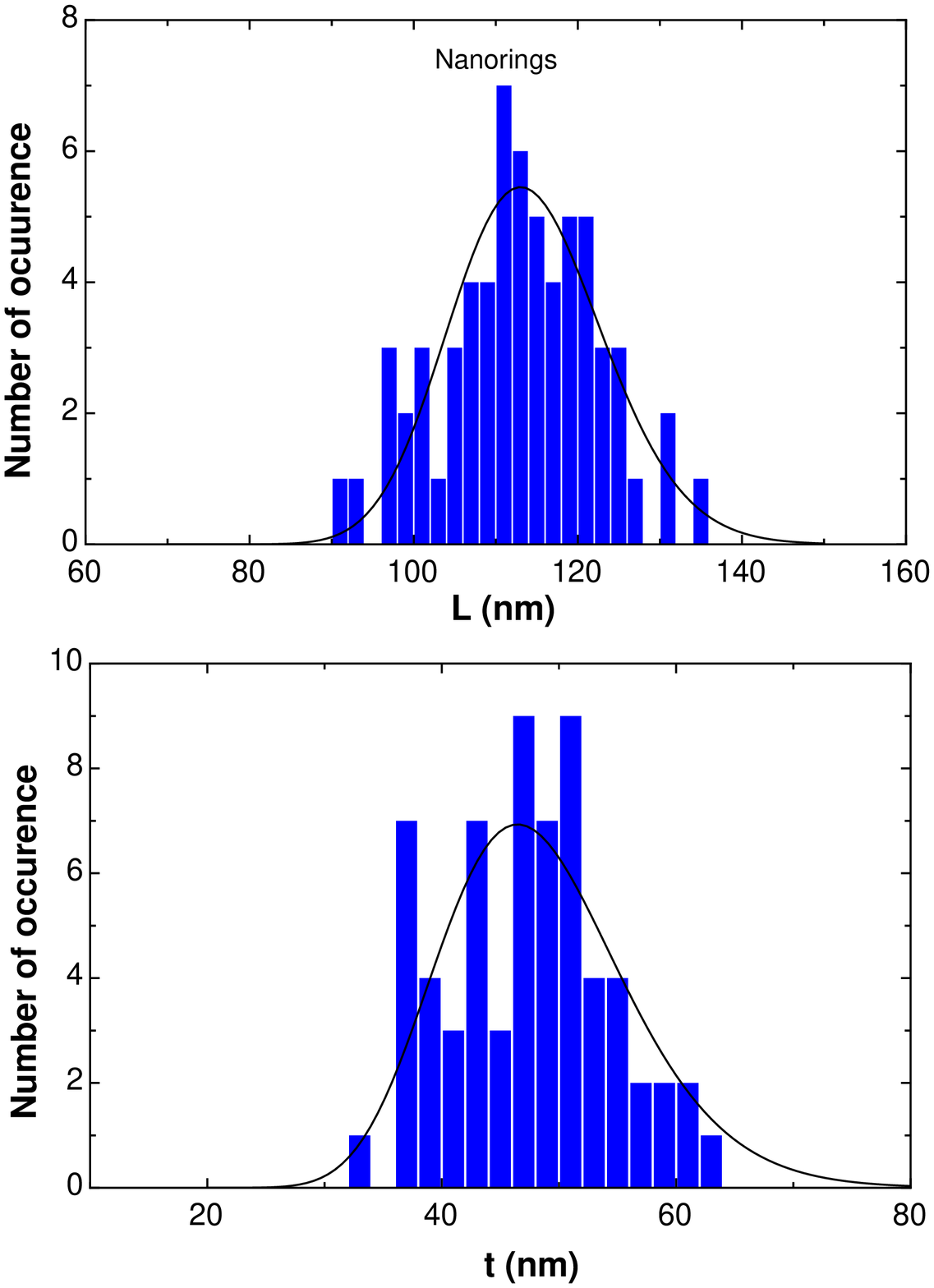}}
    \vspace{-11cm}
         \rightline{\includegraphics[height=11cm]{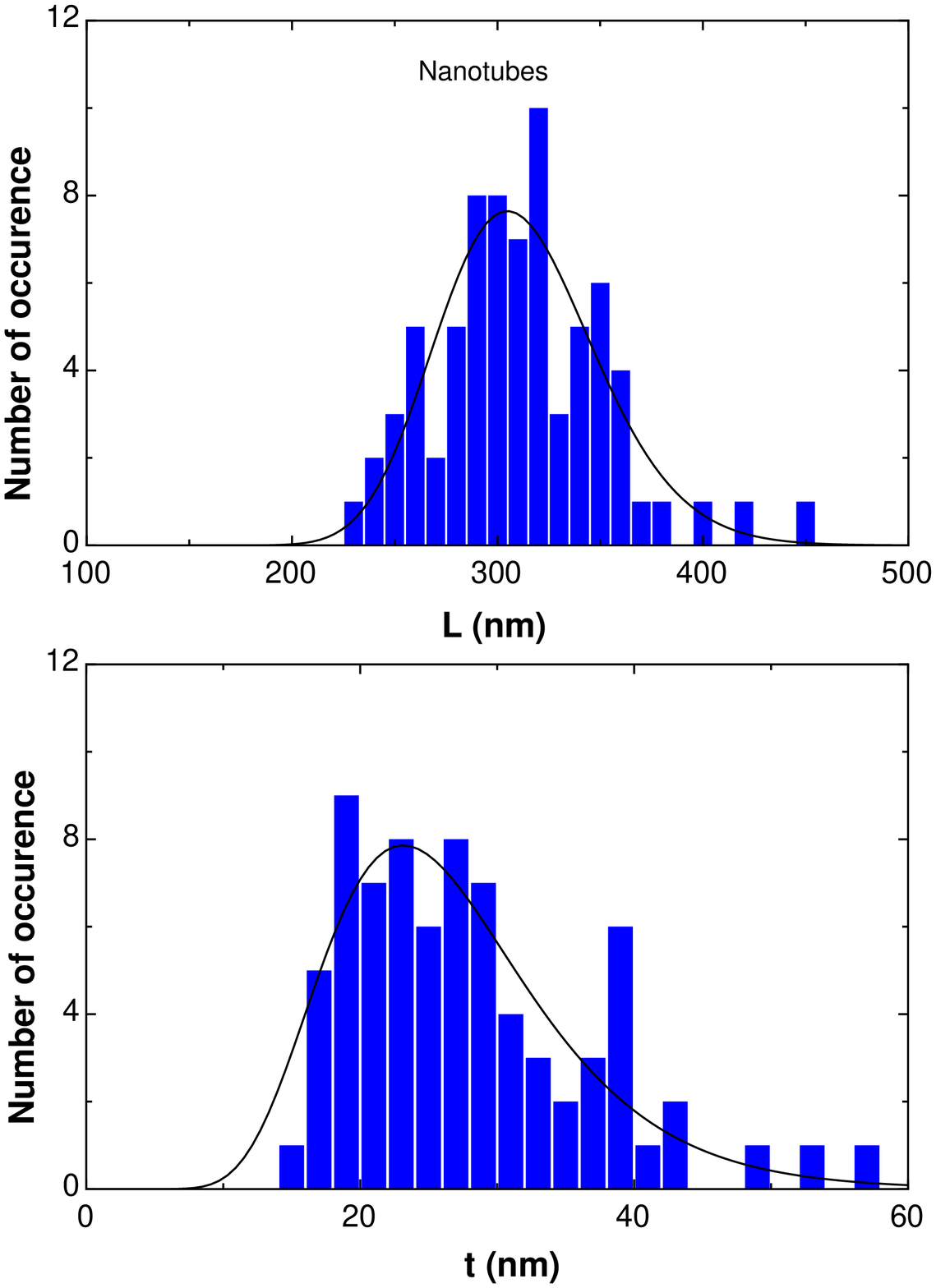}}
     \vspace{-0.1cm} 
 \caption[~]{ Length ($L$) and wall-thickness ($t$) histograms of the nanorings (left panel) and nanotubes (right panel).  The solid lines are the best fitted curves by log-normal distribution functions.  } 
\end{figure}

The transmission electron microscopic (TEM) images for the two samples are displayed in Fig.~2. The upper panel is the TEM image for the nanorings and  the lower panel is  for the nanotubes. The TEM images show much clearer morphologies of the nanocrysals than the SEM images, which allow us to obtain histograms of  their lengths and wall thicknesses. The histograms for the two samples are displayed in Fig.~3. Both length and thickness distributions are well described by a log-normal distribution function:

\begin{equation}
P(x)=\frac{1}{x\sigma\sqrt{2\pi}}\exp (-\frac{\ln^{2}(x/x_{0})}{2\sigma^{2}}),
\end{equation}

where $\sigma$ is the standard deviation and $\ln x_{0}$ is the mean value of $\ln x$.  The best fit of Eq.~1 to the data yields $L_{0}$ =113.8$\pm$0.9~nm and $t_{0}$ = 47.8$\pm$1.3~nm for the nanorings, and $L_{0}$ = 309.5$\pm$3.7~nm and $t_{0}$ = 25.6$\pm$0.9~nm for the nanotubes.  

Since x-ray diffraction intensity or magnetic moment of a particle is proportional to its volume, the mean value of  length or thickness should be length- or thickness-weighted, that is,
\begin{equation}
 x_{av} = \frac{\int_{0}^{\infty} x^{2}P(x)dx}{\int_{0}^{\infty} xP(x)}.
\end{equation}

Based on Eq.~2 and the fitting parameters for the histograms, the mean length and thickness are calculated to be 115 and 50~nm for the nanorings, respectively,  and  317 and 30~nm for the nanotubes, respectively.  The mean outer diameter of the short nanotubes are 169~nm, much larger than the mean tube length (115~nm), which is consistent with the ring morphology. The mean outer diameter of the nanotubes are 148~nm, much smaller than the mean tube length (317~nm), which is consistent with the tube morphology. It is remarkable that the mean thicknesses (50 and 30~nm) of the nanorings and nanotubes inferred from the TEM images are very close to those (58 and 32~nm) deduced from x-ray diffraction spectra (see below).

\begin{figure}[htb]
     \vspace{-0.1cm}
    \includegraphics[height=10.5cm]{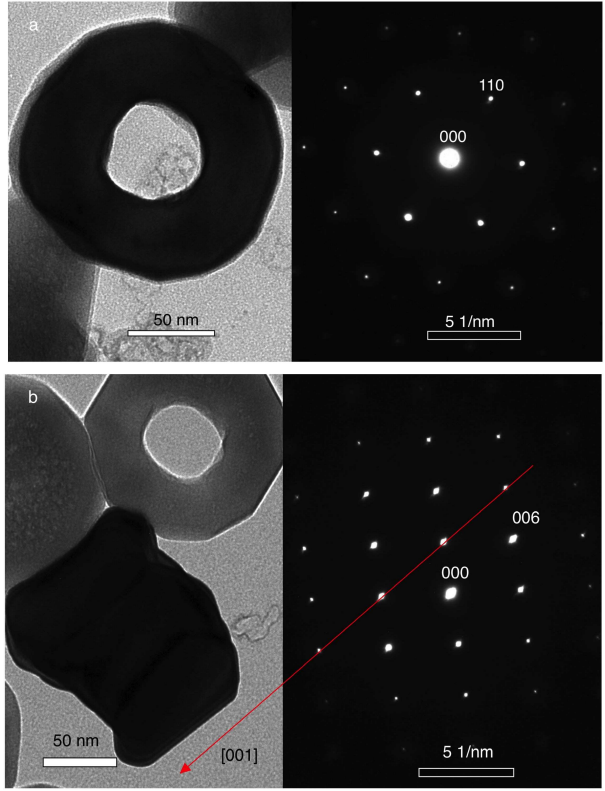}
     \vspace{-0.2cm}
 \caption[~]{TEM images  (left panels) and   SAED patterns (right panels) for a single nanoring. The results consistently demonstrate  the single-crystalline nature of the nanoring with its axis  parallel to the [001] direction. } 
\end{figure}

In the left panel of Figure 4a, we show the TEM image of a single nanoring (the top view). This ring has a wall thickness of about 50~nm, which is very close to the mean value deduced from the histogram above and slightly smaller than the mean value of 58 nm deduced from the XRD peak widths (see below). The selected-area electron diffraction (SAED) pattern (right panel of Fig.~4a) with a clear hexagonal symmetry
indicates that the nanoring is a single crystal with a ring axis parallel to the crystalline $c$-axis. In order to further prove the single-crystalline nature of the nanoring, we show the side-wall view of the ring (left panel of Fig.~4b) and the corresponding SAED pattern (right panel of Fig.~4b). The red arrow indicates the [001] direction, which is determined by the SAED pattern. It is apparent that the ring axis is parallel to the [001] direction or the crystalline $c$-axis. From the SAED pattern, we can evaluate the $c$-axis lattice constant. The obtained  $c$ = 13.77(2)~\AA~is close to that determined from the XRD data (see below).

For the nanotube sample, it is very unlikely to get a top-view TEM image since the axes of the tubes tend to be parallel to the surface of the sample substrate. So we can only take TEM images of a single nanotube from the side-wall view.    The left panel of  Fig.~5 displays a side-wall-view TEM image of a nanotube. The image indicates that the length of the tube is about 200~nm. The single-crystalline nature of the nanotube is clearly confirmed by the SAED pattern (see the right panel of Fig.~5). The red arrow marks the [001] direction, which is determined by the SAED pattern. It is striking that the tube axis is also parallel to the [001] direction. 

\begin{figure}[htb]
     \vspace{-0.1cm}
    \includegraphics[height=5.1cm]{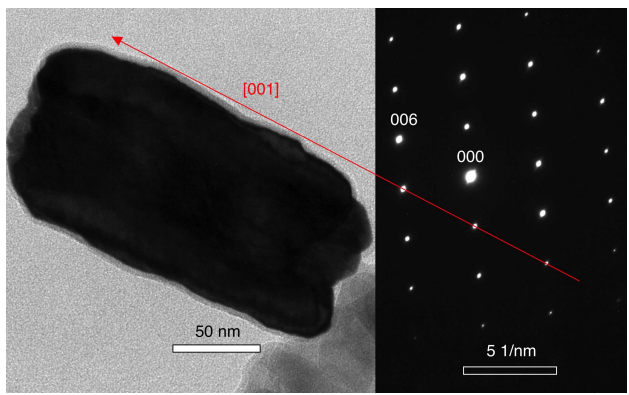}
     \vspace{-0.2cm}
 \caption[~]{TEM image  (left panel) and   SAED pattern (right panel) for a single nanotube. The results consistently demonstrate  the single-crystalline nature of the nanotube with its axis  parallel to the [001] direction. } 
\end{figure}

Figure~6 shows x-ray diffraction (XRD) spectra of two $\alpha$-Fe$_{2}$O$_{3}$ nanostructures prepared with the NH$_{4}$H$_{2}$PO$_{4}$ concentrations of 0.05 and 0.50 mM/L, respectively. The spectra were taken by Rigaku Rint D/Max-2400 X-ray diffractometer. These samples are phase pure, as the spectra do not show any traces of other phases. 
Rietveld refinement of the XRD data (see solid blue lines) with a space group of $R\bar{3}c$  (trigonal hematite lattice) was carried out to obtain the cell parameters and fractional coordinates of the atoms. The atomic occupancy was assumed to be 1.0 and not included in the refinement. We tried to include the lattice strains and particle sizes in the refinement but the uncertainties of these fitting parameters are even much larger than themselves. A large reliability factor ($R_{wp}$$\sim$11$\%$) of the refinement makes it impossible to yield reliable fitting parameters for  lattice strains (which are negligibly small) and particle sizes (which are quite large).  In contrast, the lattice parameters obtained from the refinements are quite accurate: $a$ = $b$ = 5.0311(12)~\AA, $c$ = 13.7760(33)~\AA~for the nanotube sample,  and  $a$ = $b$ = 5.0340(6)~\AA, $c$ = 13.7635(17)~\AA~for the nanoring sample. These parameters are slightly different from those for a bulk hematite \cite{Hill}: $a$ = $b$ = 5.0351~\AA, $c$ = 13.7581~\AA. 

\begin{figure}[htb]
     \vspace{-0.2cm}
    \includegraphics[height=10.5cm]{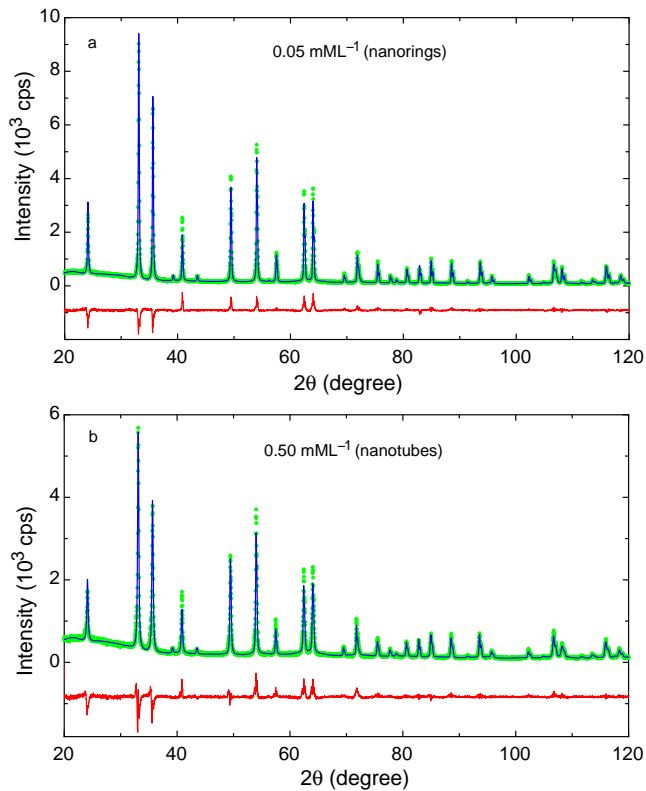}
     \vspace{-0.1cm}
 \caption[~]{X-ray diffraction (XRD) spectra of two $\alpha$-Fe$_{2}$O$_{3}$ nanostructures prepared with the NH$_{4}$H$_{2}$PO$_{4}$ concentrations of 0.05 and 0.50 mM/L, respectively. Rietveld refinement of the XRD data (solid blue lines) with a space group of $R\bar{3}c$  (trigonal hematite lattice) was carried out to obtain the lattice parameters and fractional coordinates of the atoms. The red lines represent the differences between the data and the refined curves. } 
\end{figure}

Since the axes of both nanorings and nanotubes are parallel to the crystalline $c$-axis, the mean wall thickness of the nanorings and nanotubes can be quantitatively determined by the peak widths of the  x-ray diffraction peaks that are associated with the diffraction from the planes perpendicular to the $c$-axis. Figure~7 shows x-ray diffraction spectra  of the (110), (300), and  (220) peaks for the nanoring and nanotube samples. The peaks are best fitted 
 by two Lorentzians (solid lines) contributed from the Cu $K_{\alpha 1}$ and $K_{\alpha 2}$ radiations. The fit has a constraint that the ratio of the $K_{\alpha 1}$ and $K_{\alpha 2}$ intensities is always equal to 2.0, the same as that used in Rietveld refinement.

\begin{figure}[htb]
     \vspace{-0.2cm}
   \leftline{\includegraphics[height=11cm]{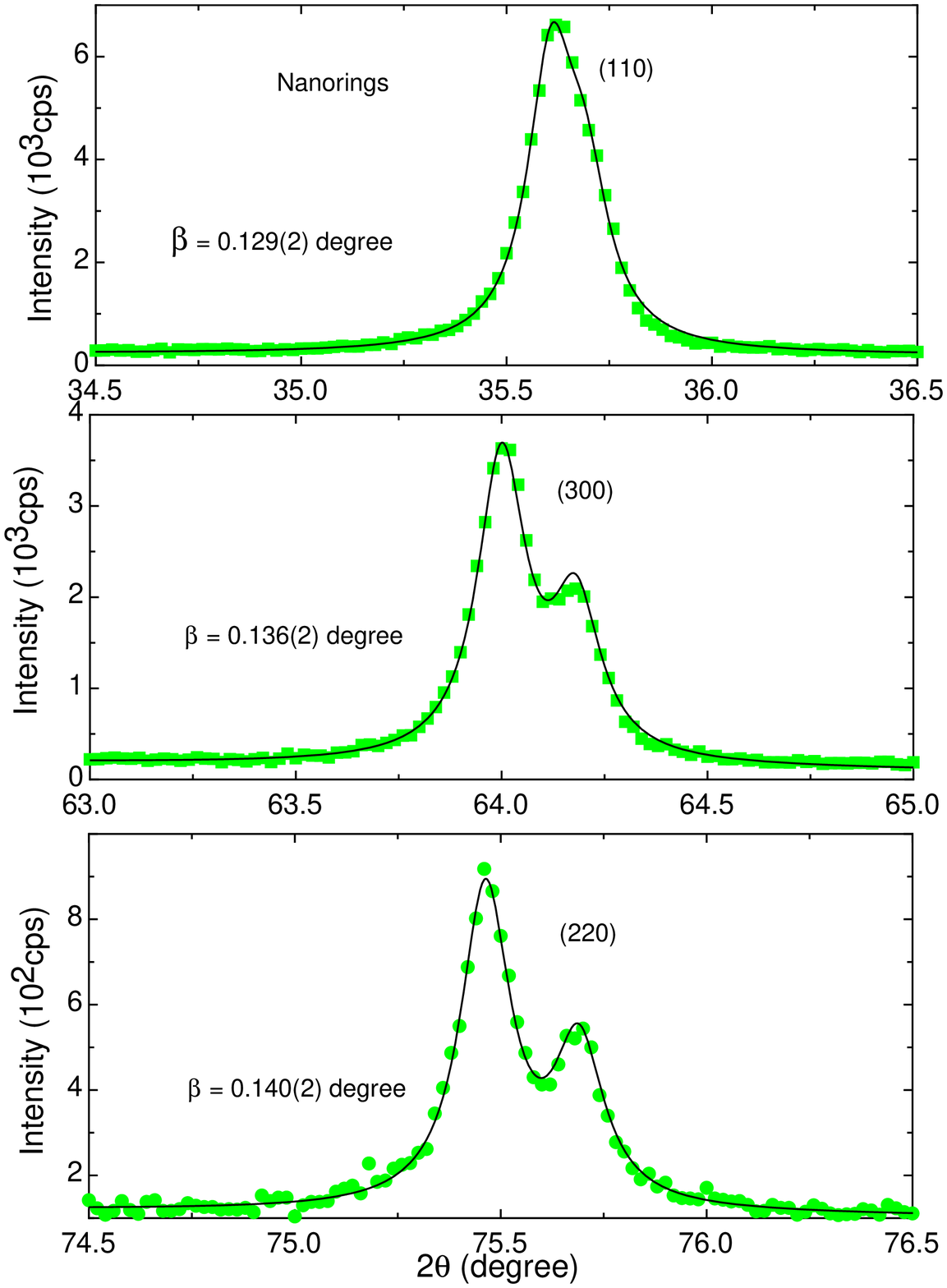}}
    \vspace{-11cm}
         \rightline{\includegraphics[height=11cm]{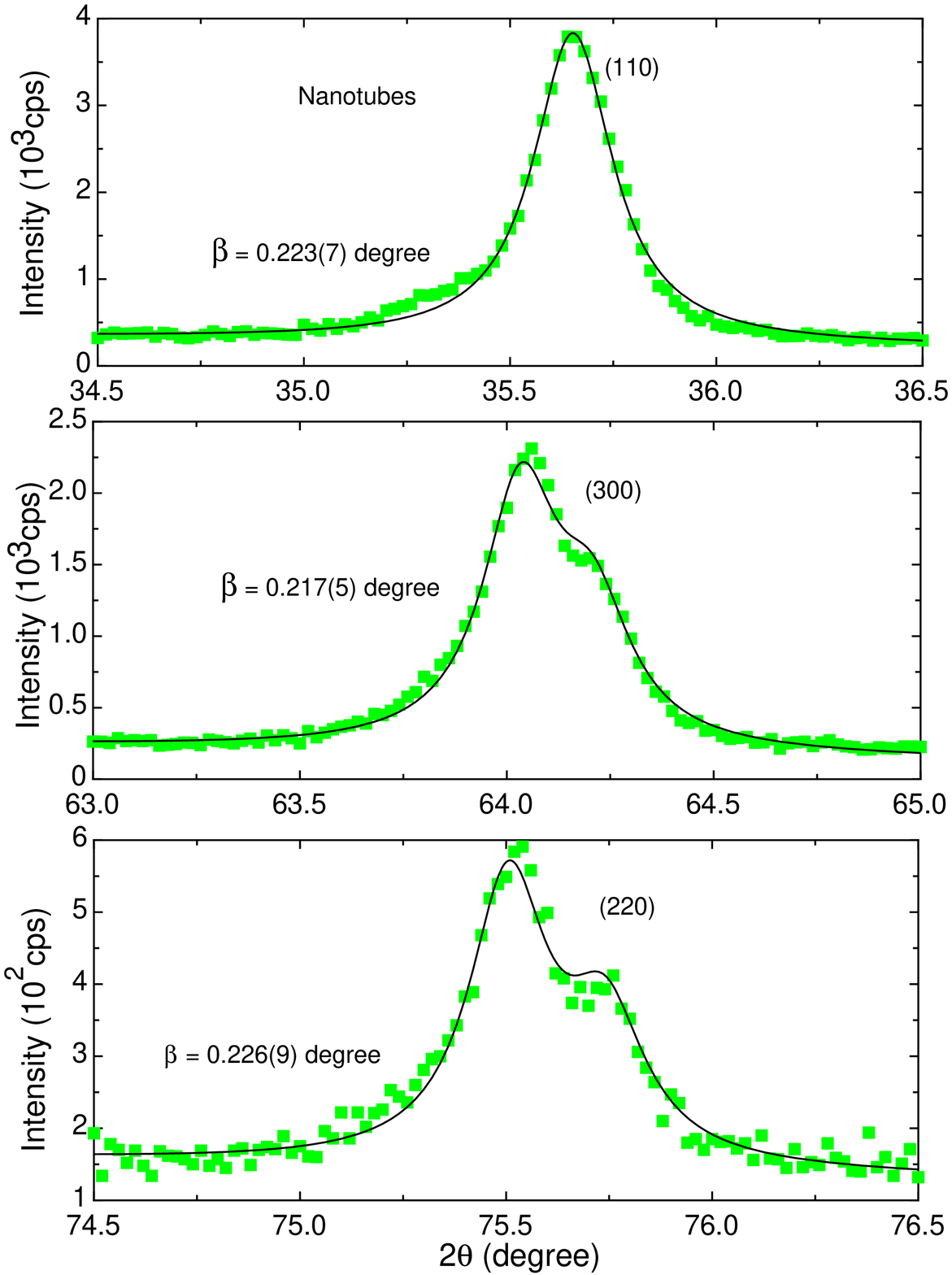}}
     \vspace{-0.1cm} 
 \caption[~]{X-ray diffraction spectra of the (110), (300), and  (220) peaks for the nanorings (left panel) and nanotubes (right panel). The peaks are best fitted 
 by two Lorentzians (solid lines), which are contributed from the Cu $K_{\alpha 1}$ and $K_{\alpha 2}$ radiations. The intrinsic peak width $\beta$ (after correcting for the instrumental broadening) is indicated in each figure.  } 
\end{figure}

It is known that the  x-ray diffraction peaks are broadened by strain, lattice deficiencies, and small particle size. When the density of lattice deficiencies is negligibly small, the broadening is contributed from both strain $\epsilon$  and  particle size $t_{av}$. In this case, there is a simple expression \cite{Hall}:
 \begin{equation}
 \frac{\beta\cos\theta}{\lambda} = \frac{0.89}{t_{av}} + \frac{2\xi\sin\theta}{\lambda},
 \end{equation}
where the first term is the same as Scherrer's equation that is related to the particle size $t_{av}$,  the second term is due to strain broadening, and $\xi$ was found to be close to 2$\epsilon$ (Ref.~\cite{Smith}).  In Fig.~8, we plot $\beta\cos\theta/\lambda$ versus  $2\sin\theta/\lambda$ for the nanorings and nanotubes. According to Eq.~3, a linear fit to the data gives information about the mean wall thickness $t_{av}$ and strain $\epsilon_{a}$ along $a$ and $b$ axes. The strain is small and negative for both samples (see the numbers indicated in the figures). It is interesting that the magnitudes of the strain inferred from the XRD peak widths  are very close to those calculated directly from the measured lattice parameters. For example, the strain is calculated to be $-$0.023(13)$\%$ from the lattice parameters for the nanorings and for the bulk, in excellent agreement with that ($-$0.017(2)$\%$) inferred from the XRD peak widths.  Moreover, the mean wall thicknesses inferred from the XRD peak widths are very close to those determined from TEM images. This further justifies the validity of our Williamson-Hall analysis of the XRD peak widths.

 \begin{figure}[htb]
     \vspace{-0.2cm}
 \includegraphics[height=10.5cm]{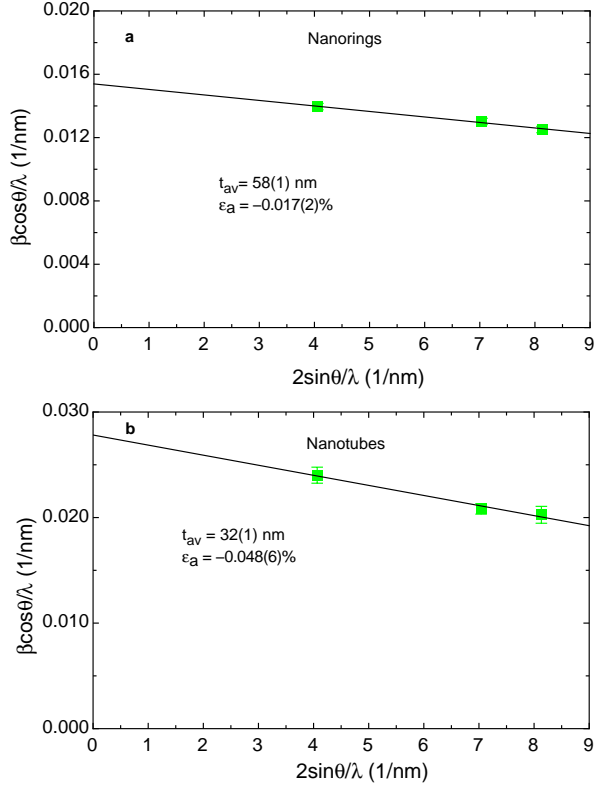}
     \vspace{-0.1cm} 
 \caption[~]{ Dependence of $\beta\cos\theta/\lambda$ on $2\sin\theta/\lambda$ for the nanotings and nanotubes. The linear lines are the fitted curves by Eq.~3. The fitting parameters are indicated in each figure. The error bars in (a) are inside the symbols and not visible.}
\end{figure}

Figure~9 shows temperature and field dependences of the normalized magnetizations 
$M(T)/M(350K)$ for the $\alpha$-Fe$_{2}$O$_{3}$  singe-crystalline nanorings and nanotubes. Magnetic moment was measured using 
a Quantum Design vibrating sample magnetometer (VSM) with a resolution better than 1$\times$10$^{-6}$ emu. The samples were initially cooled to 10~K in zero field and  a field of 100 Oe was set at 10 K, and then the moment was taken upon warming up to 350~K and cooling down from 350~K to 10~K. At 10~K, other higher fields (1 kOe, 10 kOe, and 50 kOe) were set and the moment was taken upon warming up to 350~K and cooling down from 350~K to 10~K.   
It is remarkable  that the magnetic behaviors of the two nanostructures are very different. For the nanorings,  the magnetization shows rapid  increase around 200 K upon warming, which 
is associated with the Morin transition (see Fig.~9a). It is worth noting that there seem to be two transitions with slightly different Morin transition temperatures. The reason for this is unclear. 

\begin{figure}[htb]
     \vspace{-0.2cm}
    \includegraphics[height=11cm]{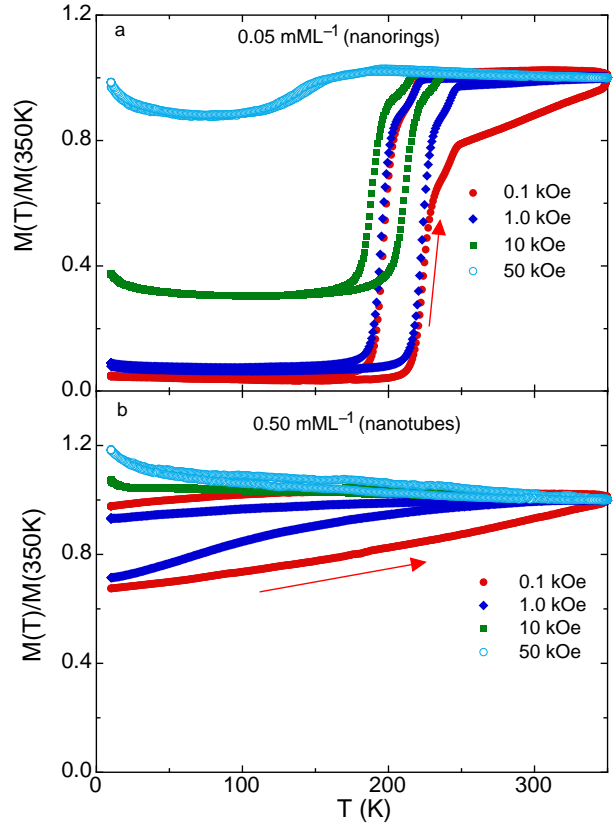}
     \vspace{-0.2cm}
 \caption[~]{Temperature and field dependencies of the normalized magnetizations 
 $M(T)/M(350K)$ for the nanorings (a) and nanotubes (b). For the nanotubes, the normalized magnetization measured in a lower field (10 Oe) is very similar to that measured in 100~Oe, suggesting absence of the Morin transition down to this low field. }  
\end{figure}

Furthermore, the magnetization below the Morin transition 
temperature $T_{M}$ is small (AF state) and it enhances significantly above 
$T_{M}$ (WF state).  It is interesting that  $T_{M}$ for heating measurements is 
significantly higher than that for cooling measurements (the arrows in the figure indicate 
the directions of the measurements). This difference is far larger than a difference 
(about 6 K) due to extrinsic thermal lag. This thermal hysteresis was also observed in spherical $\alpha$-Fe$_{2}$O$_{3}$ nanoparticles \cite{Mu}. The observed intrinsic thermal hysteresis shows that the 
nature of the Morin  transition is of  
first-order.  The result in Fig.~9a also suggests that the Morin transition temperature  decreases with the increase of the applied magnetic field. The zero-field $T_{M}$  in the nanorings is about 210~K.  What is striking is that the Morin transition is almost completely suppressed in the nanotubes (see Fig.~9b).

\begin{figure}[htb]
     \vspace{-0.2cm}
    \includegraphics[height=12.0cm]{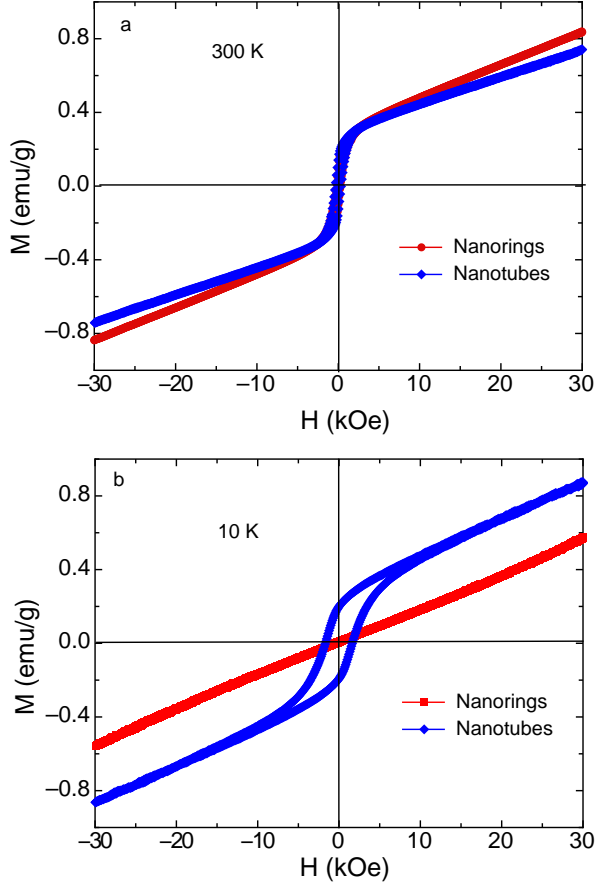}
     \vspace{-0.2cm}
 \caption[~]{Magnetic hysteresis loops at 300~K and 10~K for the nanorings and nanotubes. The room-temperature saturation magnetization $M_{s}$, as inferred from a linear fit to the magnetization data above 15 kOe, is the same (0.303$\pm$0.001 emu/g) for both samples. The saturation magnetizations for both samples are nearly the same as that (0.29$\pm$0.02 emu/g) \cite{Bod} for a polycrystalline sample with a mean grain size of about 3 $\mu$m. } 
\end{figure}

In Figure 10a, we compare magnetic hysteresis loops at 300~K for the nanorings and nanotubes. There is a subtle difference in the magnetic hysteresis loops of the two samples. The remanent magnetization $M_{r}$ for the nanotube sample is about 40$\%$ higher than that for the nanoring sample, which is related to a higher coercive field in the former sample. However, the saturation magnetization $M_{s}$, as inferred from a linear fit to the magnetization data above 15 kOe,  is the same (0.303$\pm$0.001 emu/g) for both samples. The saturation magnetizations for both samples are also the same as that (0.29$\pm$0.02 emu/g) \cite{Bod} for a polycrystalline sample with a mean grain size of about 3~$\mu$m. Since the saturation magnetization is very sensitive to the occupancy of the Fe$^{3+}$ site, the fact that the saturation magnetizations of both nanaoring and nanotube samples are nearly the same as the bulk value suggests that the occupancies of the Fe$^{3+}$ site in the nanostructural samples are very close to 1.0, which justifies our XRD Rietveld refinement. Fig.~10b shows magnetic hysteresis loops at 10~K for the two samples. It is clear that the nanotube sample remains weak ferromagnetic at 10~K (the absence of the Morin transition down to 10~K) while the nanoring sample is antiferromagnetic with zero saturation magnetization.

\section{ Discussion}

The completely different magnetic behaviors observed in the nanoring and nanotube samples are intriguing considering the fact that the two samples have the same saturation magnetization at 300~K and nearly the same lattice parameters. It is known that the lattice strain can suppress  $T_{M}$ according to an empirical relation deduced for spherical
nanoparticles \cite{Mu}:  $\Delta T_{M} = - 600\epsilon$~K, where $\epsilon$ is 
 isotropic lattice strain in $\%$. For a uniaxial strain, the formula may be modified as $\Delta T_{M} = - 200\epsilon_{i}$~K, where $\epsilon_{i}$ is the strain along certain crystalline axis. For the nanoring sample, $a$ = 5.0340(6)~\AA, which is slightly smaller than (5.0351~\AA) for a bulk hematite \cite{Hill}. This implies that $\epsilon_{a}$ = -0.023(13)$\%$  for the nanoring sample, in excellent agreement with that (-0.017(2)$\%$) inferred from the XRD peak widths. For the nanotube sample, $a$ = 5.0311(12)~\AA, so $\epsilon_{a} = -$0.080(24)$\%$, in good agreement with that (-0.048(6)$\%$) inferred from the XRD peak widths. The negative strain would imply an increase in $T_{M}$ according to the argument presented in Ref.~\cite{Schr}. Therefore the suppression of $T_{M}$ cannot arise from the lattice strains along the $a$ and $b$ directions. On the other hand, the lattice strain along the $c$ direction is positive. Comparing the measured $c$-axis lattice parameters of the two samples with that for a bulk hematite \cite{Hill},  we can readily calculate that  $\epsilon_{c}$ = 0.040(12)$\%$ for the nanoring sample and 0.130(24)$\%$ for the nanotube sample. This would lead to the suppression of $T_{M}$ by 8(2)~K and 26(5)~K for the nanoring and nanotube samples, respectively. The small negative strains along $a$ and $b$ directions are compensated by the positive strain along $c$ direction (also see Table II below) so that the volume of unit cell remains nearly unchanged. This implies that  the $T_{M}$ suppression due to lattice strains should be negligibly small.

 As mentioned above, there is also an independent  finite-size effect on $T_{M}$ unrelated to the strain. For spherical nanoparticles,   $T_{M}$ is suppressed according to $\Delta T_{M} = -1300/d$~K (Ref.~\cite{Mu}), where $d$ is the mean diameter of spherical particles in nm. For the nanoring and nanotube samples, the smallest dimension is the wall thickness, which should play a similar role as the diameter of spherical particles \cite{ZhaoPRB}. With $t_{av}$ = 58~nm and 32~nm for the nanoring and nanotube samples, respectively,  the suppression of $T_{M}$ is calculated to be 22~K and 41~K, respectively. Therefore, due  to the finite-size effect,  $T_{M}$ would be reduced from the bulk value of 258~K (Ref.~\cite{Schr}) to 236~K and 217~K for the nanoring and nanotube samples, respectively. For the nanoring sample, the zero-field $T_{M}$  is about 211~K, which is 25~K lower than the expected value from the finite-size effect only. This additional $T_{M}$ suppression of 25~K should be caused by other mechanism discussed below. For the nanotube sample,  the Morin transition is almost completely suppressed, which cannot be explained by the strain and/or finite-size effect.

    \begin{figure}[htb]
     \vspace{-0.2cm}
    \includegraphics[height=7cm]{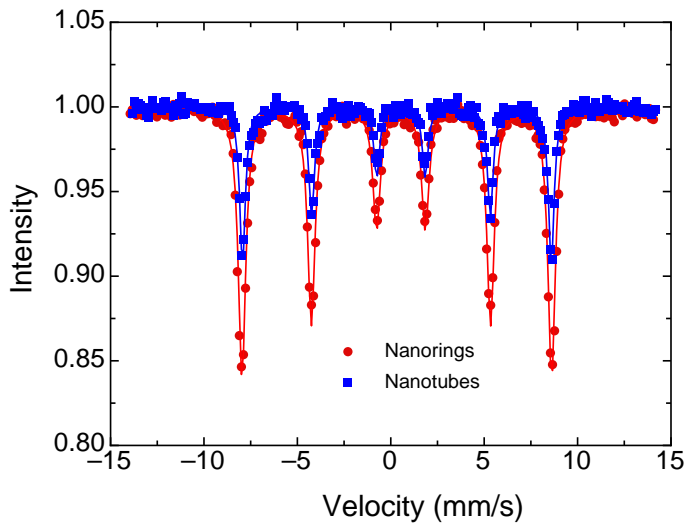}
     \vspace{-0.2cm}
 \caption[~]{Room-temperature M\"ossbauer spectra for the nanorings and nanotubes. The spectra are fitted by a single
sextet (solid lines) with the fitting parameters displayed in Table I. } 
\end{figure}

Another possibility is that the nanotubes may contain more lattice deficiencies than the nanorings.  If this were true, the line width of the M\"ossbauer spectrum for the nanotube sample would be broader than that for the nanoring sample because the line width is sensitive to disorder, inhomogeneity, and lattice deficiencies. In contrast, the observed line width for the nanotube sample is smaller than that for the nanoring sample by 33$\%$ (see Fig.~11 and Table I).  If there would exist substantial lattice deficiencies, they would mostly be present in surface layers.  The narrower M\"ossbauer line width observed in the nanotube sample is consistent with the fact that the nanotubes have a smaller fraction of surface layers.   Moreover, the room-temperature M\"ossbauer spectra of both samples show only one set of sextet, suggesting no superparamagnetic relaxation at room temperature. This is consistent with the observed magnetic hysteresis loops (see Fig.~10a).

 \begin{table}[htp]
 \caption[~]{The fitting parameters for the room-temperature M\"ossbauer spectra of the nanorings and nanotubes}
\begin{center}
    \begin{tabular}{ | l | l | l | l | p{3cm} |}
    \hline
     & Half width (mm/s)& Hyperfine field (kOe) & Isomer shifts (mm/s) &Quadrupole shifts (mm/s)\\ \hline
    Nanorings & 0.240$\pm$0.005 & 513.30$\pm$0.20 & 0.44$\pm$0.01& -0.220$\pm$0.005\\ \hline
    Nanotubes & 0.180$\pm$0.005 & 511.85$\pm$0.23 & 0.44$\pm$0.01&-0.220$\pm$0.005\\ \hline
    \end{tabular}
\end{center}
 \end{table}

 Finally, we can quantitatively explain the strong shape dependence of the Morin transition temperature if we assume that the surface magnetic anisotropy constant $K_{s}$ is negative in the surface planes parallel to the $c$-axis and positive in the surface planes perpendicular to the $c$-axis. Indeed a negative value of $K_{s}$ was found in Ni (111) surface \cite{Grad} while $K_{s}$  is positive in Co(0001) surface \cite{Chap}.  For the nanorings, the surface area for the planes parallel to the $c$-axis are similar to that for the planes perpendicular to the $c$-axis. Therefore, the total $K_{s}$ will have a small positive or negative value due to a partial cancellation of the $K_{s}$ values (with opposite signs) in different surface planes. In contrast,  the surface area of a nanotube for the surface planes parallel to the $c$-axis  is much larger than that for the planes perpendicular to the $c$-axis. This implies that the total $K_{s}$ in the nanotubes should have a large negative value.  

For bulk hematite, the Morin transition temperature is uniquely determined by the total bulk anisotropy constant $K$ at zero temperature \cite{Art}. Contributions to $K$ are mainly  dipolar anisotropy constant  $K_{MD}$, arising from magnetic dipolar interaction,  and fine structure anisotropy (magneto-crystalline anisotropy) $K_{FS}$, arising from spin-orbit coupling \cite{Art}. With $K_{MD}$ = -9.2$\times$10$^{6}$ erg/cm$^{3}$ and $K_{FS}$= 9.4$\times$10$^{6}$ erg/cm$^{3}$ in the bulk hematite, the Morin transition temperature was predicted to be 0.281$T_{N}$ = 270~K (Ref.~\cite{Art}), very close to the measured bulk value of 258~K (Ref.~\cite{Schr}).

\begin{figure}[htb]
     \vspace{-0.2cm}
    \includegraphics[height=6cm]{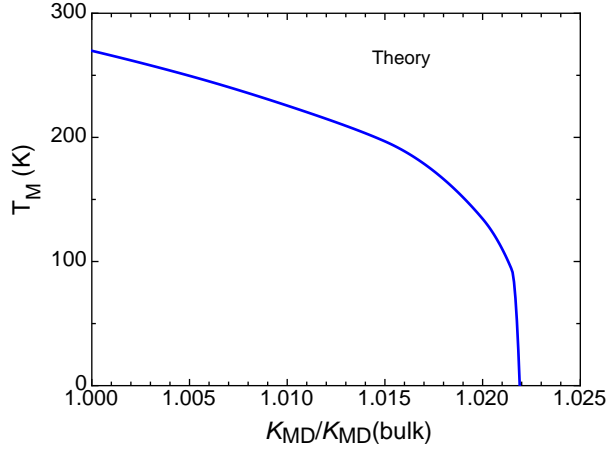}
     \vspace{-0.2cm}
 \caption[~]{ Numerically calculated $T_{M}$ as a function of $K_{MD}$/$K_{MD}$(bulk), where $K_{MD}$(bulk) is the bulk anisotropy constant. The calculation is based on a simple model presented in Ref.~\cite{Art} and on the assumption that $K_{FS}$ remains unchanged. } 
\end{figure}

Following this simple model, we can numerically calculate $T_{M}$ as a function of $K_{MD}$/$K_{MD}$(bulk) on the assumption that  $K_{FS}$ remains unchanged, where $K_{MD}$(bulk) is the bulk anisotropy constant. The calculated result is shown in Fig.~12. It is apparent that $T_{M}$ is suppressed to zero when the magnitude of $K_{MD}$ increases by 2.2$\%$. Near this critical point,  $T_{M}$ decreases rapidly with increasing the magnitude of $K_{MD}$.  For nanotubes, contribution of the surface anisotropy is substantial and should be added to the total anisotropy constant.  Following  the expressions used in Refs.~\cite{Grad,Chap}, we have 
\begin{equation} 
K_{MD} = K_{MD}(bulk) + 2K_{s\parallel}/t_{av} + 2K_{s\perp}/L_{av}, 
\end{equation}

where $K_{s\parallel}$ and $K_{s\perp}$ is the surface anisotropy constants for the planes parallel and perpendicular to the $c$-axis, respectively,  and $L_{av}$ is the average tube length. Here we have assumed that the surface areas of the inner and outer walls are the same for simplicity.  For the nanotubes, $T_{M}$ is nearly suppressed to zero. According to Fig.~12,  $K_{MD}$/$K_{MD}$(bulk) should be close to 1.022 for the nanotubes with $L_{av}$ = 317~nm and $t_{av}$ = 32~nm.  For the nanorings (short nanotubes), the zero-field $T_{M}$ is about 211~K. This implies that $T_{M}$ is totally suppressed by 47~K compared with the bulk value of 258~K. Since the finite-size effect can suppress $T_{M}$ by 22~K (see discussion above), the additional suppression of $T_{M}$ by 25~K should be due to an increase in $K_{MD}$ by about 0.58$\%$ according to Fig.~12, that is, $K_{MD}$/$K_{MD}$(bulk) =1.0058 for the short nanotubes with  $L_{av}$ = 115~nm and $t_{av}$ = 58~nm.  Substituting these $K_{MD}$/$K_{MD}$(bulk), $L_{av}$, and  $t_{av}$ values  of both nanoring and nanotube  samples into Eq.~4, we obtain two equations with two unknown variables, $K_{s\parallel}$ and $K_{s\perp}$. Solving the two equations for the unknown $K_{s\parallel}$ and $K_{s\perp}$ yields $K_{s\parallel}$ = -0.37~erg/cm$^{2}$ and $K_{s\perp}$ = 0.42~erg/cm$^{2}$. The deduced magnitudes of the  surface anisotropy constants are in the same order of  the experimental values found for  Ni and Co  [$K_{s}$ = -0.22 erg/cm$^{2}$ for Ni(111) and 0.5 erg/cm$^{2}$ for Co(0001)] \cite{Grad,Chap,Bru}. Therefore, the observed intriguing experimental results can be naturally explained by a negative and a positive surface anisotropy constant in the surface planes parallel and perpendicular to the crystalline $c$-axis, respectively. 

 \begin{table}[htp]
 \caption[~]{Some parameters for six nanostructures. The parameters for  ellipsoidal, spindle,  flattened, and rhombohedral structures are calculated from the data reported in Ref.~\cite{Mitra}. The Fe  occupancies for the nanoring and nanotube structures are inferred from the measured saturation magnetizations (see discussion in the text).  }
\begin{center}
    \begin{tabular}{ | l | l | l | l | p{3cm} |}
    \hline
     & $T_{M}$ (K) & $\epsilon$ ($\%$) &Fe occupancy& $c/a$\\ \hline
     Ellipsoidal & 251.4 & 0.028(1) & 0.9600(3)& 2.7337(1)\\ \hline
    Spindle & 245.4  & 0.063(1)&0.9921(10)&2.7329(1)\\ \hline
    Flattened & 231.5 & -0.132(2) & 0.9787(3)& 2.7327(1)\\ \hline
    Rhombohedral  &220.8  & 0.053(2)& 0.9934(16)&2.7340(1)\\ \hline
    Nanoring & 211 & -0.007(25)& 1.0& 2.7341(5)\\ \hline
    Nanotube & $<$10 & -0.028(48) & 1.0&2.7381(9)\\ \hline
    \end{tabular}
\end{center}
 \end{table}

Now we discuss the shape dependence of the Morin transition temperature observed in other nanostructures \cite{Mitra}. It was shown that $T_{M}$ shifted from highest 251.4~K for ellipsoidal to lowest 220.8~K for rhombohedral structure, with intermediate values of $T_{M}$ for the other two structures. In Table II, we compare some parameters for four nanostructures reported in Ref.~\cite{Mitra} and two nanostructures reported here. The total lattice strain $\epsilon$ is calculated using $\epsilon = 2\epsilon_{a} + \epsilon_{c}$, where $\epsilon_{i}$ is the percentage difference in the lattice constant of a nanostructure and the bulk.  It is apparent that $T_{M}$ does not correlate with any of these parameters.   For example, the Fe occupancy (0.96) in the ellipsoidal structure is significantly lower than 1.0, but  $T_{M}$ is the highest and close to the bulk value. This implies that the Fe vacancies should have little effect on the Morin transition. We thus believe that the weak shape dependence of the Morin transition observed in the previous work \cite{Mitra} should also arise from the opposite signs of the surface magnetic anisotropic constants.  The much lower $T_{M}$ in the rhombohedral structures can be explained as due to a much larger surface area parallel to the $c$ axis in the structure, in agreement with the observed HRTEM image \cite{Mitra}.
 
\section{Conclusion}

 In summary, we have prepared single-crystalline hematite  nanorings and nanotubes using  a hydrothermal method. High-resolution transmission electron microscope  and selected-area electron diffraction confirm that the axial directions of both nanorings and nanotubes  are parallel to the crystalline $c$-axis. Magnetic measurements show that
there exists a first-order Morin transition at about 210 K in
the nanoring crystals while this transition disappears in nanotube crystals.  The current results suggest that the Morin
transition depends very strongly on the shape of
nanostructures. This strong shape dependence of the Morin transition can be well explained by a negative and a positive surface anisotropy constant in the surface planes parallel and perpendicular to the crystalline $c$-axis, respectively.

{\bf Acknowledgment:} This work was supported by  the National Natural Science Foundation of China (11174165), the Natural Science Foundation of Ningbo (2012A610051),  and the K. C. Wong Magna Foundation.

~\\
$^{a}$ wangjun2@nbu.edu.cn~\\
$^{b}$ gzhao2@calstatela.edu

\bibliographystyle{prsty}

\end{document}